\documentclass[
     ,draft            
]
 {aipproc}

\layoutstyle{6x9}
\newcommand{\gaga}{$\gamma \gamma$}
\newcommand{\ega}{$e \gamma$}
\newcommand{\ra}{\rightarrow}

\begin{document}

\vspace{-1.5cm}
       \begin{flushright}
       \begin{tabular}{l}
       {\small IFT - 2002/16}    \\
       {\small July 2003}
       \end{tabular}
       \end{flushright}

\title{
Photon-Photon and Electron-Photon Physics 
or Physics at Photon Collider
}

\author{Maria Krawczyk}{
  address={Institute of Theoretical Physics,Warsaw U.}
}

\begin{abstract}
 A (updated) summary of the  {Photon-Photon and Electron-Photon}  
physics session is presented.
\end{abstract}

\maketitle


\section{Introduction}
The energetic, highly polarized  photons can be 
produced at a high rate in the Compton process as a result of  the 
backscattering of bright laser-photons on high-energy electrons. 
This is a basic concept of the Photon Collider (Compton Collider) - 
called also a Photon Linear Collider (PLC) \cite{ginz-princ,telnov}. 
Such collider
is  considered as an option to be realized in the  $\gamma \gamma$ 
and $e \gamma$ modes at all discussed future 
$e^+e^-$  Liner Colliders (LC): TESLA, NLC, JLC and CLIC \cite{space}.

The potential of Photon Collider is very rich, in some cases even spectacular
\cite{ginz-princ,telnov,gg2000,badelek,acfa,bbb,golden,hagiwara,mayda}. 
First of all, PLC  
seems to be 
  the most suitable place to observe a scalar (Higgs) sector.
Especially useful should be a $\gamma \gamma$ mode, 
where the  C=+ resonances,   like  Higgs boson(s), can be produced,
and determination of their basic properties 
can be performed with a high accuracy. Such a mode allows for a
precision measurement of the fundamental effective (loop-) coupling 
 $\gamma \gamma h$, which involves contributions from all fundamental, massive 
charged particles, with   masses originating from the Higgs mechanism. 
In general, such contributions do not decouple  and therefore
this effective  coupling is sensitive to heavy particles,  even if their  
masses are in a range well 
beyond  a reach of the existing and even the next generation experiments.
PLC with highly polarized photon-beams offers excellent opportunity 
for testing of new interactions, including CP-violating one, 
in the scalar  sector. 
It should discover   heavy spin-zero particles, e.g.
Higgs bosons from extended models, supersymmetric particles, etc..., 
in the  mass range not accessible at other experiments. This is due to the fact
that at PLC such particles can be produced singly, 
in contrast to  a $e^+e^-$  LC, where they are produced in pairs 
or with other heavy particles.  
The $e\gamma$ mode allows to measure another effective (loop-)
coupling, namely  $Z\gamma h$.  
This mode  is of particular importance for studies of the anomalous $W$-
interactions in a process $e \gamma \ra \nu W$, and   for an  investigation 
of the slepton sector.
The PLC, both in the $\gamma \gamma$ and  $e\gamma$ modes, is a perfect place 
to study QCD, in particular $t$-quark interaction and a 
hadronic ``structure'' of the real photon.

The realistic simulations of the  basic processes
for various  designs of the Photon Collider  
are  being performed currently, allowing for a reliable comparison 
with similar simulations done for the main $e^+e^-$ option and for the hadron 
colliders. 

There are  specific aspects of the $\gamma \gamma$ and $e \gamma$ colliders, 
absent at $e^+e^-$ colliders, which arise  mainly from the facts that 
the photon-beams have a wide spread 
of energy  and a varying with energy degree of   polarization. 
Also, a  large hadronic background is expected at a Photon Collider, 
as  photon may (with a probability $\alpha$) behave like a hadron 
(vector meson dominance idea). To deal with all these issues  
the dedicated ``tools'' are needed. 

The photon-photon and photon-electron physics working group
has a mission to investigate a physics potential of a Photon Collider.
There is a large activity in this working group - more than 15 talks were 
presented during this workshop, most of them in the joint sessions with 
other working groups like Higgs, SUSY, EW and QCD.
A special panel discussion was organized to discuss a need of a
Photon Collider.

\section{Photon Collider}
%
The maximum energy of the back-scattered photons produced
in the Compton scattering, to be used as a beam for PLC,
 is equal to  $E_{\gamma}^{max}=\frac{x}{x+1}E_0$, where 
$x=\frac{4E_0\omega_{laser}}{m_e^2}$.  
This energy may reach up to 80 \% of the 
energy of the initial electrons $E_0$ (for $E_0=250$ GeV and laser energy 
$\omega_{laser}=$ 1.17 eV  ($\lambda=1.06 ~\mu m$) and  $x=$4.5 \cite{ginz-princ,telnov}). 
For larger $x$, the   energy spectrum for the photon-beam is  more  
monochromatic, being peaked at a high energy.
However, if  $x$ becomes  too large (in the above example larger than 4.8) then
 the  $e^+e^-$ pairs can be created
in the collision of the back-scattered photons  with the laser photons. This 
 process  leads to decrease of the luminosity of the Photon Collider. 
For $x=4.8$ 
the maximum CMS energy of the $\gamma \gamma$ collision  is equal to 
80 \% of the energy for the $e^+e^-$ collision and it reaches 
90 \%  for the $e \gamma$ case. If lower energy of Photon Collider is needed,
one can use the same laser and decrease electron  energy keeping all beam
parameters as for higher energy (a bypass
solution), then  the luminosity for $\gamma \gamma$ collider is proportional 
to $E_0$
\cite{telnov,tohru}. One  considers also a technique 
with a tripled laser frequency (NLC/JLC) \cite{tohru,asner-gunion}.  
This way even 
at smaller electron beam-energy $E_0$ $x$   remains large
and a resulting energy spectrum for the back-scattered photons is
peaked \cite{tohru,asner-gunion}.

The Photon Collider  can be realized using only $e^-$ beams.
A dedicated interaction point, with a finite crossing angle to avoid 
background from the disrupted beams, is foreseen at all LC colliders. 
If electron bunches  are tilted with respect to the direction of the 
beam motion (``crab''-crossing scheme),  the luminosity is the same as for
the  head-on collisions. 
Luminosities of the $\gamma \gamma$ and $e \gamma$ colliders  are of the 
order of 10 \% of the $e^-e^-$ geometric luminosity,  which can be larger 
than the corresponding luminosity for a $e^+e^-$ collider \cite{tohru,telnov}.

The shape (monochromaticity) of the energy-spectrum 
and the degree of polarization of the back-scattered photons 
depend crucially on the (product of) polarization of the initial electrons 
and  laser photons. 
The 80 \%  polarization for the parent electrons (i.e. 
twice the electron helicity
<$2\lambda_e$> equal to $\pm 0.8$) is  feasible at LC while 
 a circular polarization of the laser photons, $P_c$, can be close to $\pm 1 $.
When  $2\lambda_eP_c=-0.8$ one obtains a highly monochromatic and highly polarized photon-beam. It has  a characteristic high-energy peak in
 the energy spectrum for   $E_{\gamma}$ larger than 0.8 $E^{max}_{\gamma}$,
and   its
 circular polarization  reaches up to 100 \%  
for the photon energy close to $E_{\gamma}^{max}$ (the average 
polarization is equal to 90-95 \% for a high-energy part). 
Also, a linear polarization
of the photon-beams  can be obtained by using a linearly 
polarised laser-light. The maximum degree of a linear polarization 
of the back-scattered photon, <$l$>, is obtained 
at its maximum energy, unfortunately this maximum degree is higher for a lower 
$x$ (up to  63.3 \% for $x=$1.8 and only 33.4 \% at $x=$4.8 
\cite{telnov,tohru}.).

If both photon-beams are obtained from parent particles with
$2\lambda_eP_c=-0.8$,
then the luminosity spectra for  the $\gamma \gamma$ collision has a 
high-energy peak (for $z=W_{\gamma \gamma}/2 E_0$ larger that $ 0.8 ~z_{max}$),
with a width at half maximum of about 15 \% and the integrated 
luminosity  about 1/3 of the luminosity for $e^+e^-$ collider  
\cite{telnov,gg2000}, e.g. for TESLA 84 $fb^{-1}/yr$. In the considered case 
both   photon-beams have a 
high degree of circular polarization, <$\lambda$>,<$\lambda'$> $\approx -P_c$, 
with $P_c =1 $, hence     <$\lambda \lambda'$>  is close to 1. 
This means a domination of  a state with a projection of the total angular 
momentum on the $z$-axis, $J_z=\lambda -\lambda'$, equal to 0.

By flipping the helicities and polarization of the parent electrons and laser 
photons for one or both photon-beams, one can easily  optimize the
Photon Collider to work either as a factory or a discovery machine. 
In particular, to produce in a $\gamma \gamma$ mode  a particle with a 
definite mass $m$, one  rather chooses a factory design, i.e. with the   
monochromatic spectra ( $2\lambda_eP_c=-0.8$, possibly large $x$), 
to tune a maximum  of $W_{\gamma \gamma}$
to the mass $m$ (however first  an energy of the $ee$ collision  has to be 
adjusted: $\sqrt {s_{ee}} \approx m$/0.8). A high rate for the spin-zero 
state is an extra benefit of this option.
On other hand, a maximum degree of linear polarization of the photon-beam,
 transferred from a linearly polarised laser-photons, 
is high enough to select in an effective way  states with a definite CP-quantum
number,
i.e.  CP-even or CP-odd states,  
depending whether 
the  polarization-vectors of colliding photons are parallel or perpendicular.
To  discover new particles, a discovery design of PLC with 
 the  broad energy spectra of photon-beams (e.g. with $2 \lambda_eP_c=+0.8$) 
will be more useful. 

\section{Tools}  
\begin{figure}
  \includegraphics[height=.3\textheight]{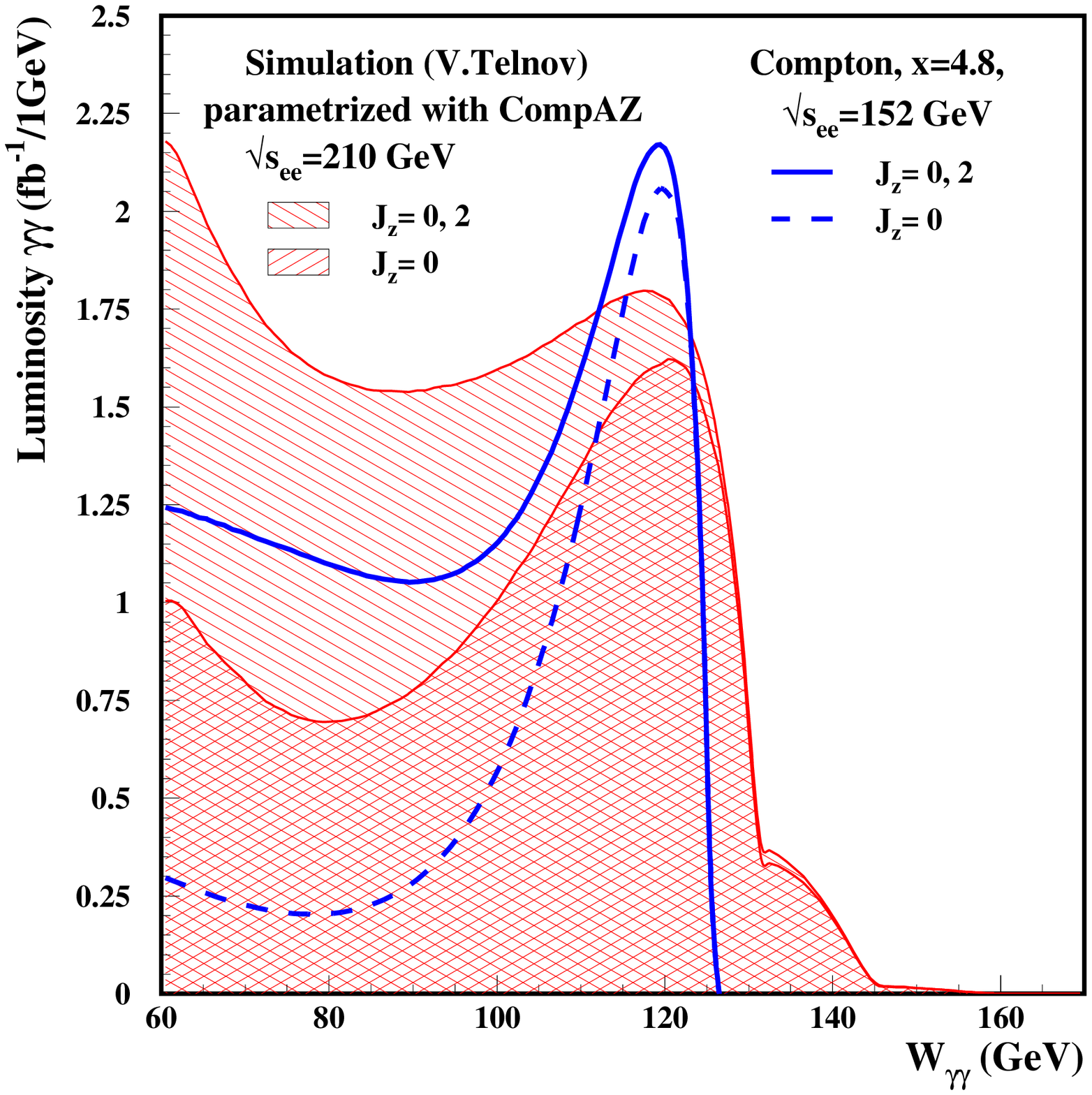}
\includegraphics[height=.3\textheight]{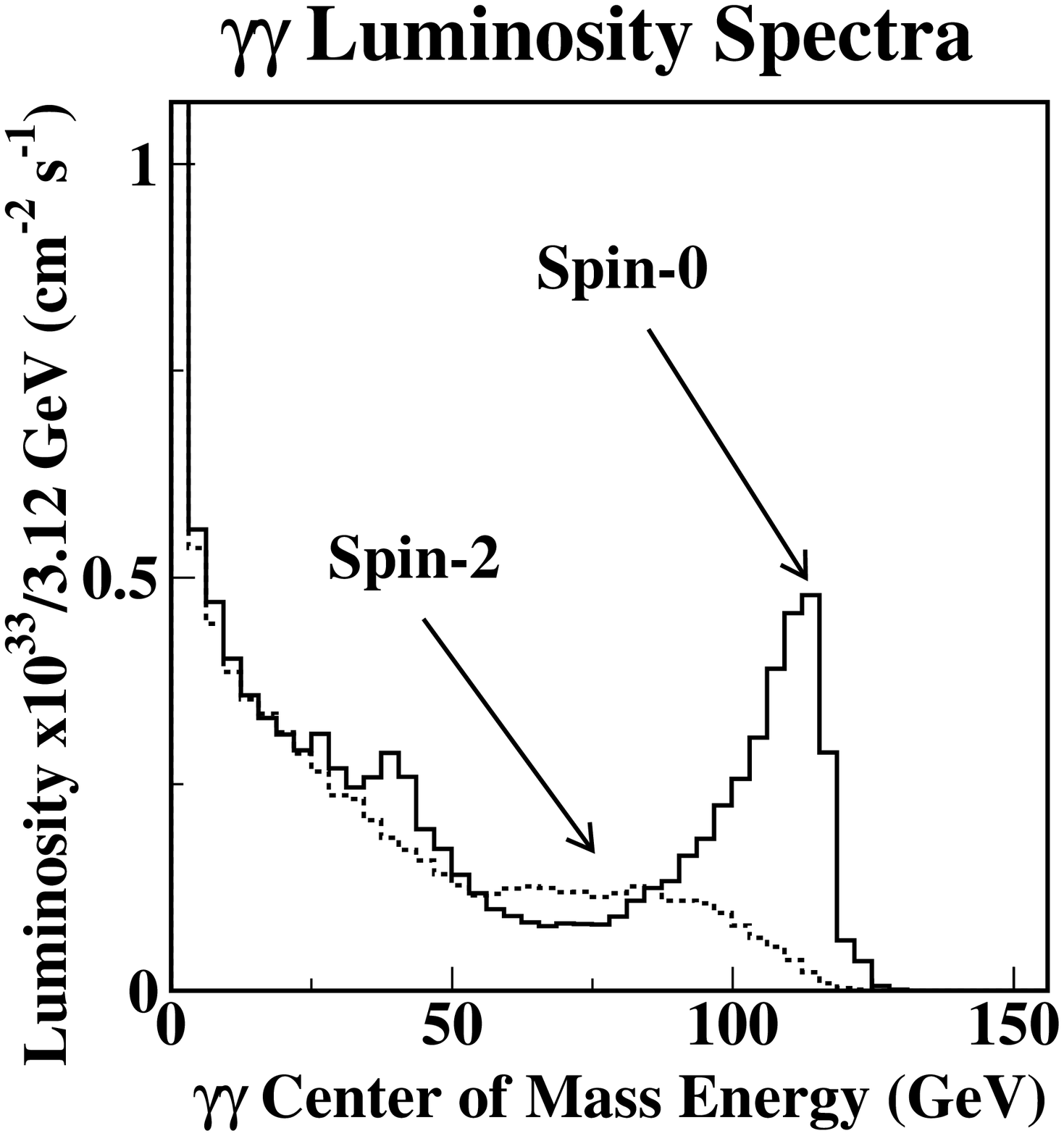}
\includegraphics[height=.3\textheight]{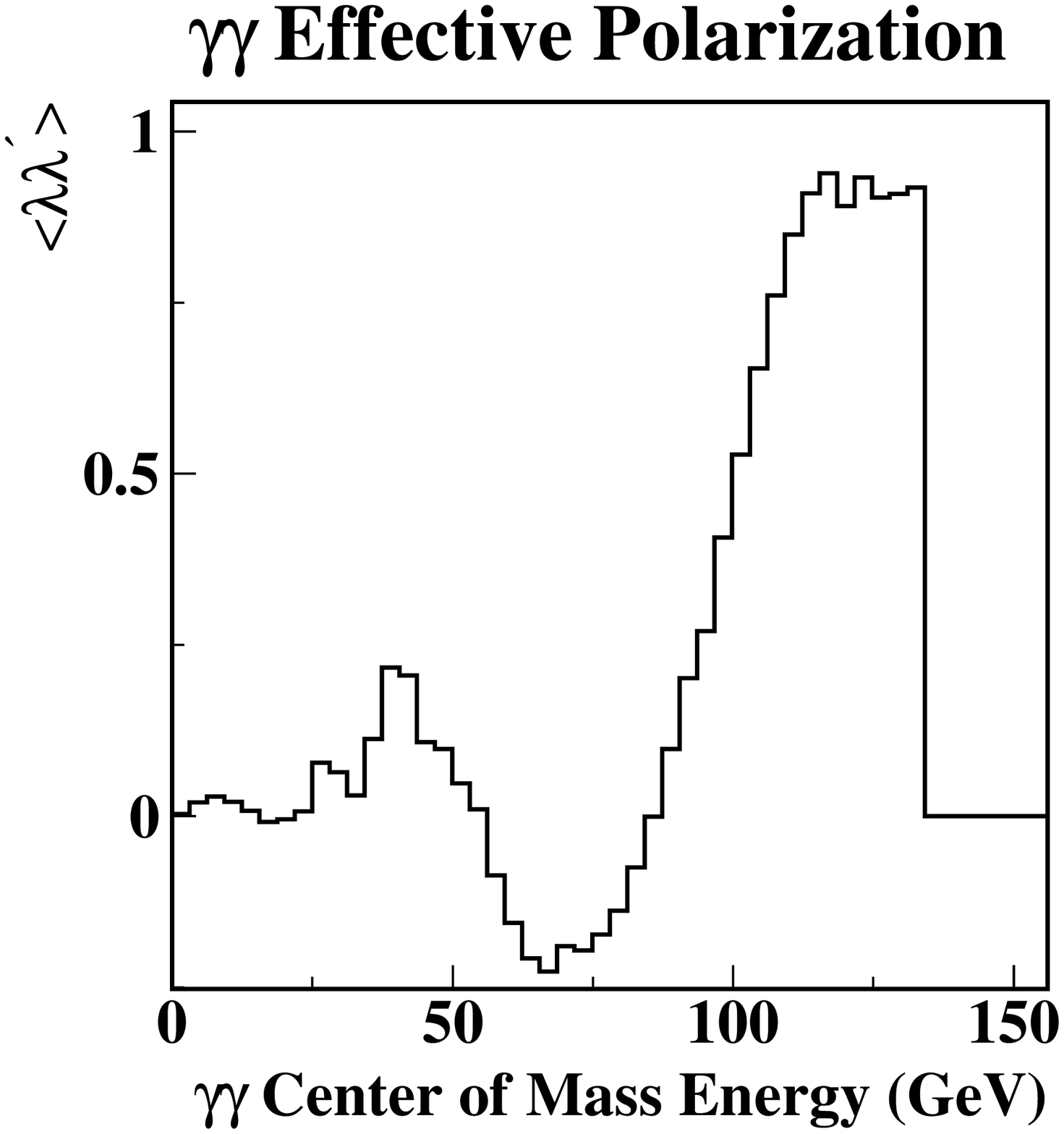}
  \caption{ The peaked spectra considered for a production of a Higgs boson 
with mass 120 GeV for $2 \lambda_e P_c=-0.8$. Left: The realistic ($\sqrt{s_{ee}}=$ 210 GeV, $x=1.89$)  and ideal 
Compton ($\sqrt{s_{ee}}=$ 152 GeV, $x=4.8$) luminosity
spectrum for TESLA, for $J_z=$0,$\pm 2$ 
\cite{nzk-b}; Midle: The luminosity 
spectra for NLC
($\sqrt s_{ee}=$150 GeV, $x=4.1$) for $J_z=$0,$\pm 2$;
Right: The avarage polarization <$\lambda \lambda'$> for  NLC, parameters as in (Midle), from   \cite{mayda}.}
\label{fig1}
\end{figure}

The realistic spectra for the photon-beam, which include
the  higher order QED processes, differ significantly from an  
ideal (the lowest-order) Compton form.  The programmes  generating the 
realistic luminosity spectra  for the 
$\gamma \gamma$ and $e \gamma$ colliders  exist for all machines 
 (CAIN and PHOCOL with CIRCE and  an analytic  CompAZ parametrization
\cite{cainphocolcircecompaz}).
The luminosity 
spectra 
for the TESLA (both  an ideal Compton and a realistic (Telnov) spectrum) 
and for NLC, are  presented in Fig.~\ref{fig1}(Left) and  
Fig.~\ref{fig1}(Midle). They
correspond to the product of 
helicities of the initial particles for each photon-beam equal to
$2\lambda_eP_c=-0.8$. The energy of $e^-e^-$ is tunned to give
a high-energy peak for the invariant mass $W_{\gamma \gamma}$ equal to 120 
GeV (equal to e.g. a Higgs boson mass). 
The individual contributions, for  $J_z$ equal to 0 and $\pm 2$,
 are shown. Also the average product of helicities of the back-scattered 
photons <$\lambda \lambda '$>  for a  NLC $\gamma \gamma$ collider is shown
in the  Fig.~\ref{fig1} (Right).

Existing generators like PYTHIA, and 
GRACE, PANDORA, CompHep \cite{MC} are being used by various groups to simulate 
events  in the $\gamma  \gamma$  and $e \gamma$ collisions;  
proper matching of the 
matrix elements for basic hard processes and  fragmentation/hadronization 
processes, as planned e.g. in the program APACIC++\cite{frank}, is needed. 

For a realistic simulation of the  photon-initiated processes  
 at PLC,  we need to model properly the 
hadronic interaction of photons. Recently,  a new parton 
parametrization for the real photon was constructed \cite{cjkl}. It uses  
a full set of available data for $F_2^{\gamma}$ and is based on 
ACOT$_{\chi}$ scheme. Such scheme,  implemented for a first time
for a photon, offers an improved treatment of the 
heavy-quark contributions.  A  proper description of 
heavy-quark production in a resolved-photon processes at PLC is nessesary
to make  a reliable estimation of the signal and  background in a Higgs-boson 
search. Of a great importance is also a detailed
study of the total cross-section for $\gamma \gamma \rightarrow hadrons$,
 presented during this workshop \cite{adr}.

%
\section{Golden processes}
%
There are significant differences between  the basic processes which may 
appear in   $e^+e^-$ and in  $\gamma \gamma$ collisions \cite{bbb,golden}. 
At a fixed energy of the corresponding collision, the production rates for 
a particular state with pair of scalars, fermions or vector particles 
are larger for the $\gamma \gamma$  than the $e^+e^-$ case. 
In the $\gamma \gamma$ collision 
a resonant production of the  $J^{PC}=0^{++},0^{-+},2^{++},...$ states may 
occur \cite{lan}, in contrast to the $e^+e^-$ case, with the s-channel resonances
$J^{PC}=1^{-+},... $.
With a opportunity to produce  a zero-spin $C=+$ resonance,  
a  Photon Collider can be treated as a Higgs factory. In this case,
a Higgs boson should be  found at other collider, and  knowing  its mass 
one can adjust the energy of the collider using monochromatic spectra,
as shown  in  Figs.~\ref{fig1}. It is 
fortunate, that at the same time, the $J_z=0$ state can be produced at a high 
rate. This   enhances a signal while suppresses some of important background
processes, e.g. $\gamma \gamma \rightarrow b\bar b$ \cite{golden,ssr}. 
Higgs factory gives an opportunity for the precision 
measurements of  the mass,  spin, parity, and the CP-nature of the Higgs boson 
from SM and beyond \cite{gg2000,golden,hagiwara}. 
A Photon Collider can   serve  also as a discovery machine for heavier Higgs 
bosons or other new particles -
a broad energy spectrum is preferred then \cite{telnov,bbb,golden,asner-gunion}. The advantage of a 
Photon Collider is that here one can  search for a Higgs boson, and study
its properties, up to a higher mass  than 
at the  $e^+e^-$ collider, since it can be produced 
singly at PLC in contrast to main discovery channels at $e^+e^-$ colliders. 
In addition, a
 Photon Collider offers  an opportunity to search for a medium mass Higgs boson,  with small or zero coupling to ZZ or WW  (as H and A in a decoupling 
scenario of MSSM), which are not  accessible at the $e^+e^-$ LC option, see
discussion below.  

The fundamental quantity to measure at Photon Collider is the Higgs 
decay-width 
$\Gamma_{\gamma \gamma}$, and also $\Gamma_{Z \gamma}$,
which is  sensitive to all fundamental, massive  charged particles of the 
underlying theory with masses from the Higgs mechanism \cite{shifman}. Both decay widths
 can be   measured with a high precision,
especially the two-photon width, what allows to distinguish various 
extensions of the SM, even in their
decoupling limits or SM-like scenarios, what we  discuss below. If the 
measurements of 
these decay widths are combined with results of  measurements 
of the corresponding branching ratios at the  $e^+e^-$ LC, the total
decay width for Higgs particle can be derived with accuracy dominated by
the expected error of Br($h \ra \gamma \gamma$), of the order of  10 \%
\cite{tohru96,ssr,nzk-b,nzk-b-ams}.

The W-boson production in the $\gamma \gamma$ and  $e \gamma$ options of 
a Photon Collider is  sensitive to the anomalous gauge 
couplings \cite{golden,bbb,moenig}. Photon Collider also allows to perform  
the dedicated QCD studies,
among them  on  top-quark physics  and  on the  ``structure'' of a real-photon
\cite{rindani}.
Also, a search for  new particles, e.g.
SUSY particles, are  considered as an   unique opportunity
of the  Photon Collider. New interactions, among other the  NC QED, 
models with higher dimensions, and a Higgs-radion mixing 
(Randall-Sundrum model), etc.
 can be tested at
PLC with a high precision \cite{gg2000,godfrey,mayda}.

%
\subsection{SM and SM-like Higgs boson}
In light of LEP data a light Higgs (with mass above 115 and below 200 GeV)
 is expected in the Standard Model.
The  SM-like scenarios in which one light scalar exists with the 
basic (tree-level) 
couplings as predicted in SM, and all other Higgs particles 
with masses larger than $\sim$ 800 GeV,
may be realized in many models with an extended Higgs sector. In particular, in
2HDM  or MSSM one can consider a  limit of a very 
large $M_A$ or $M_{H^+}$. Yet some  deviations, i.e. non-decoupling effects,  
may  appear in the 2HDM, see e.g. \cite{espiru,gko-ga,gko,hab,kanemura}.
This can show up in the loop-couplings, 
such as $\gamma \gamma h$ or $Z\gamma h$, due to the  
additional contributions of  the charged Higgs boson,
and/or  other charged particles from the extended models.

A light SM (or SM-like) Higgs boson with mass below $\sim$ 140 GeV  
decays dominantly   to  the $b\bar b$. 
With a proper $b$-tagging and  after correcting for the escaping neutrinos,
 one can achieve precision of a 
measurement of the $\Gamma_{\gamma \gamma}\cdot $Br$(H\rightarrow b\bar b)$
at TESLA  
at the level 1.6 \% (Fig.2 (Left)) \cite{nzk-b}. 
For higher Higgs masses the precision 
for of such  measurement 
worsens, it goes up to 7 \% for mass equal to 160 GeV, see Fig.\ref{fig3} 
(Left) \cite{nzk-b-ams}. 

\begin{figure}
\includegraphics[height=.3\textheight]{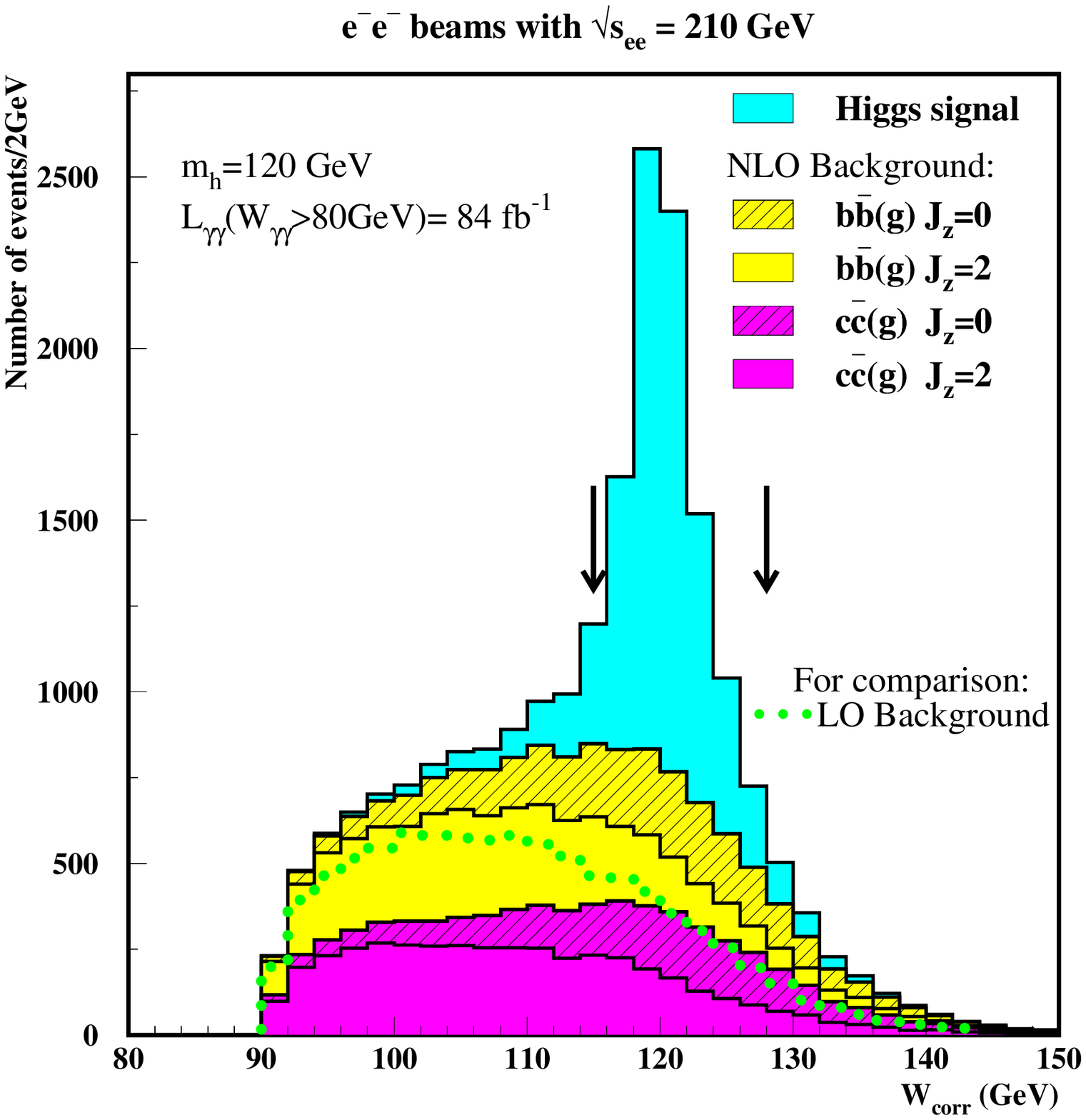}
\includegraphics[height=.3\textheight]{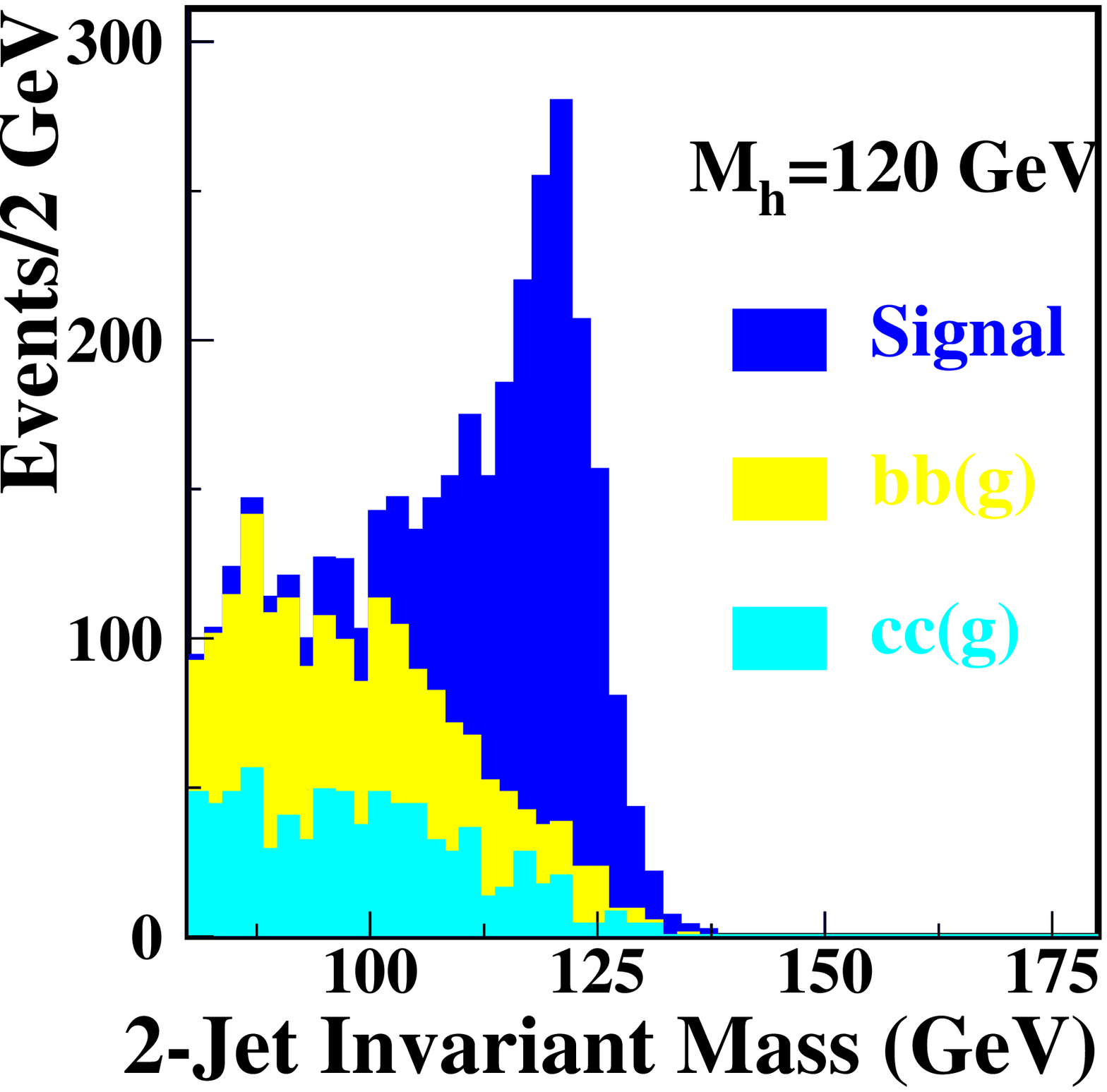}
\caption{The SM Higgs resonance at PLC for a peaked spectra (with
$2 \lambda_e P_c=-0.8$); Left: for TESLA, 
with NLO background estimation, $b$-tagging, and corrections for 
escaping  neutrino ($\sqrt{s_{ee}}=210$ GeV, $x=$ 1.8), from 
\cite{nzk-b};  Right: for NLC,
with LO background ($\sqrt{s_{ee}}= 160$ GeV, $x$=4.334) \cite{asner-gunion}.}
\label{fig2}
\end{figure}

For heavier Higgs bosons the WW and ZZ (also  W$^*$ and Z$^*$) decays 
 channels become important in the SM and SM-like scenarios.
 The interference between the signal and the (SM-) background, which is very 
large for the WW channel, 
has to be taken into account \cite{cidwaj,nzk-w}. 
The   Fig. \ref{fig3} (Left) shows a  comparison  of the 
accuracy of measuring 
of $\Gamma_{\gamma \gamma}$   for the $b \bar b$ and WW plus ZZ final-states.
The yellow line corresponds to the deviation from the SM prediction
due to the  contribution of 
a charged Higgs boson $H^+$, with mass 800 MeV, expected  in the SM-like 
limit of the 2HDM II \cite{gko-ga}.
As it  shown in Fig. ~\ref{fig3} (Right),  the measurements of the partial 
width  and of the phase of the amplitude give complementary  
 information and provide a tool with a  strong discriminating power 
between SM and various SM-like scenarios. In the figure
the results for the SM-like models with one extra heavy particle 
with mass 800 GeV ($H^+$,  quark U or D, lepton L) are shown. 

\begin{figure}
\includegraphics[height=.3\textheight]{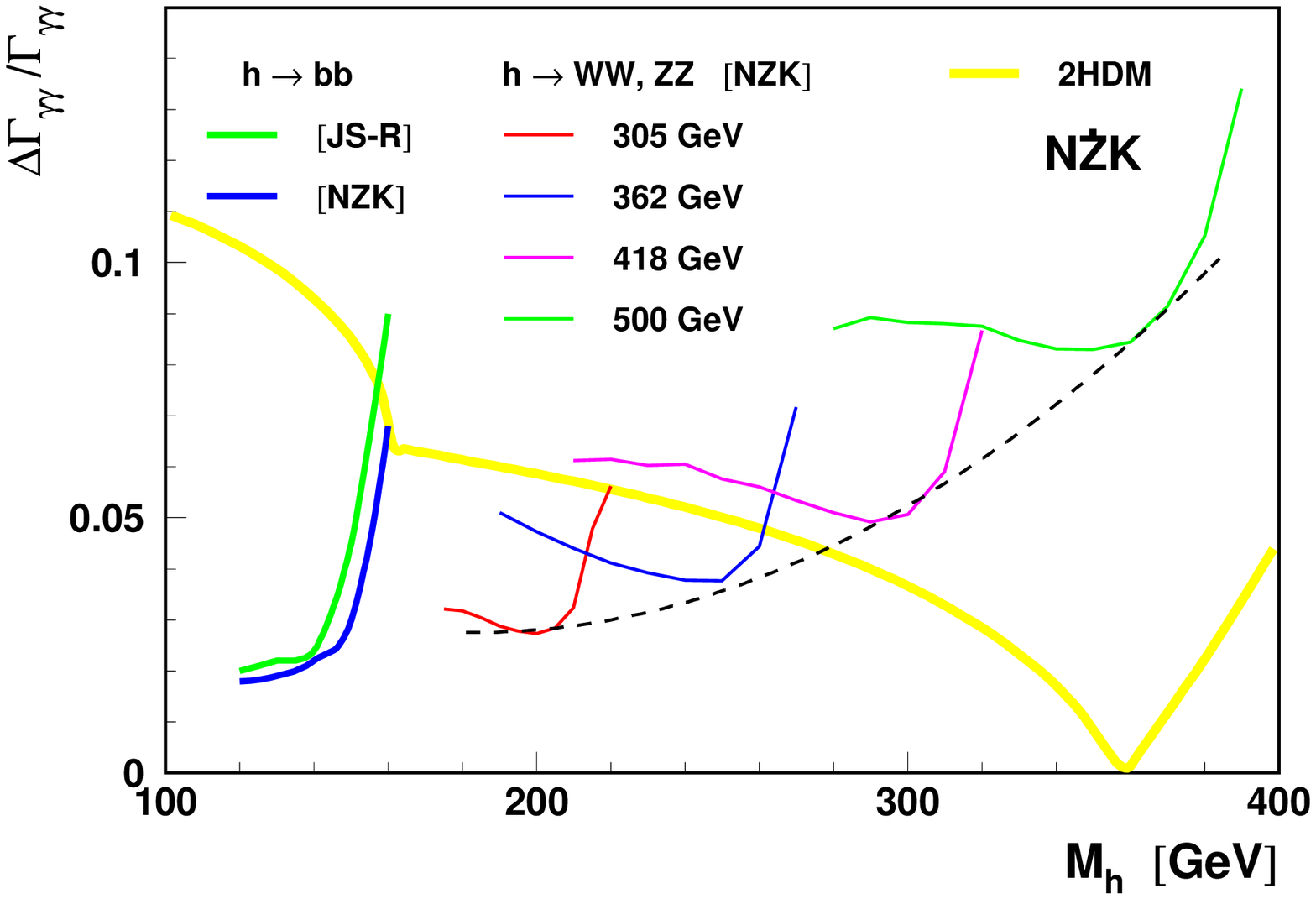}
\includegraphics[height=.3\textheight]{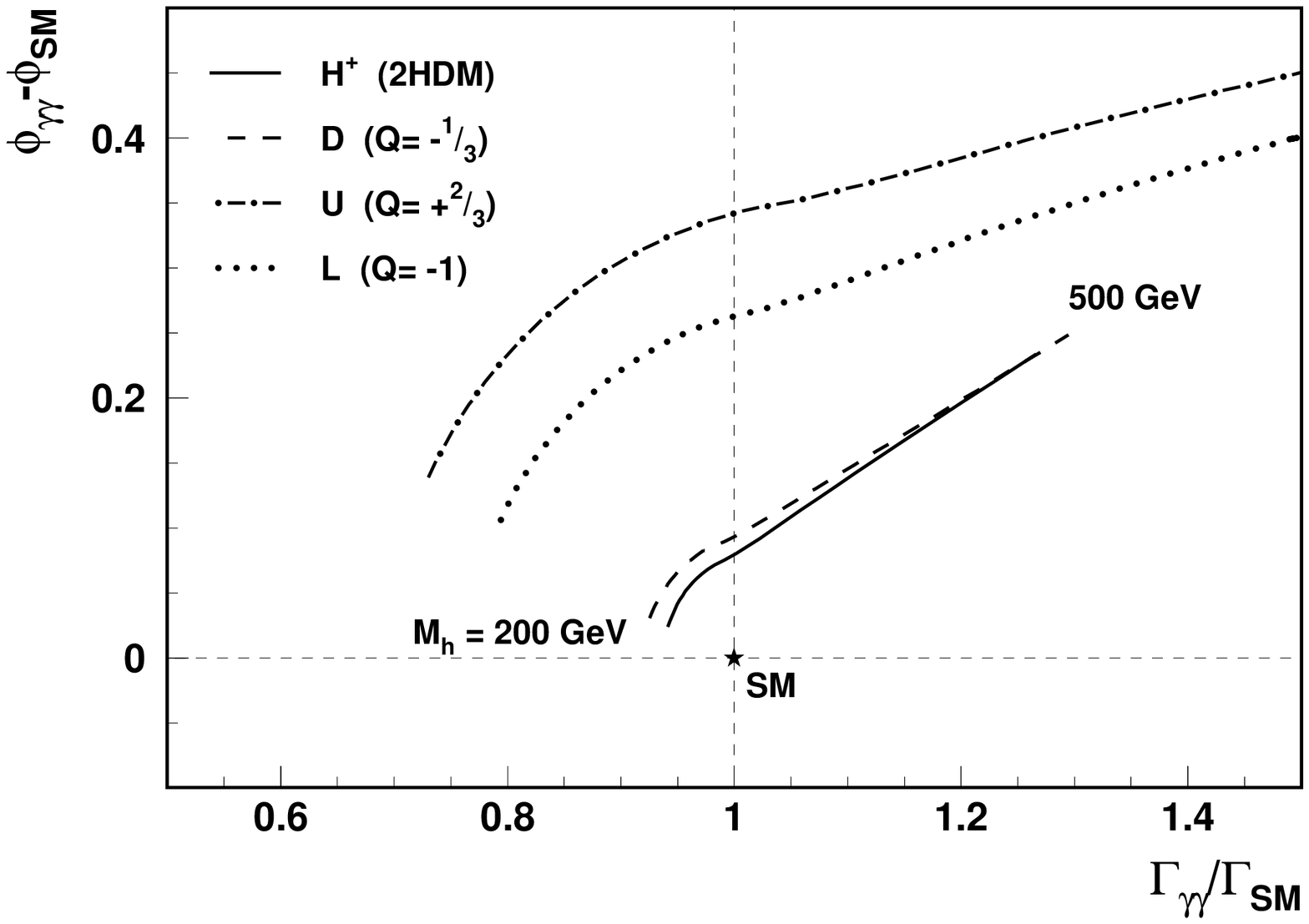}
\caption{Left: The accuracy of the measurements of the 
partial decay width  $\Gamma_{\gamma \gamma}$ 
as a function of mass of the SM Higgs boson $h$  (decays into  $b \bar b$
\cite{nzk-b-ams} 
and $WW/ZZ$ \cite{nzk-w}). The deviation for a possible  SM-like 2HDM II 
with the charged Higgs boson with mass 800 GeV
($\mu=0$) is also shown; Right:
Deviation from SM of the  phase of amplitude and of $\Gamma_{\gamma \gamma}$, 
for various mass of $h$ and various SM-like models with 
one extra heavy particle with mass equal to 800 GeV, from 
\cite{nzk-w}.}
\label{fig3}
\end{figure}
%

\begin{figure}
\includegraphics[height=.3\textheight]{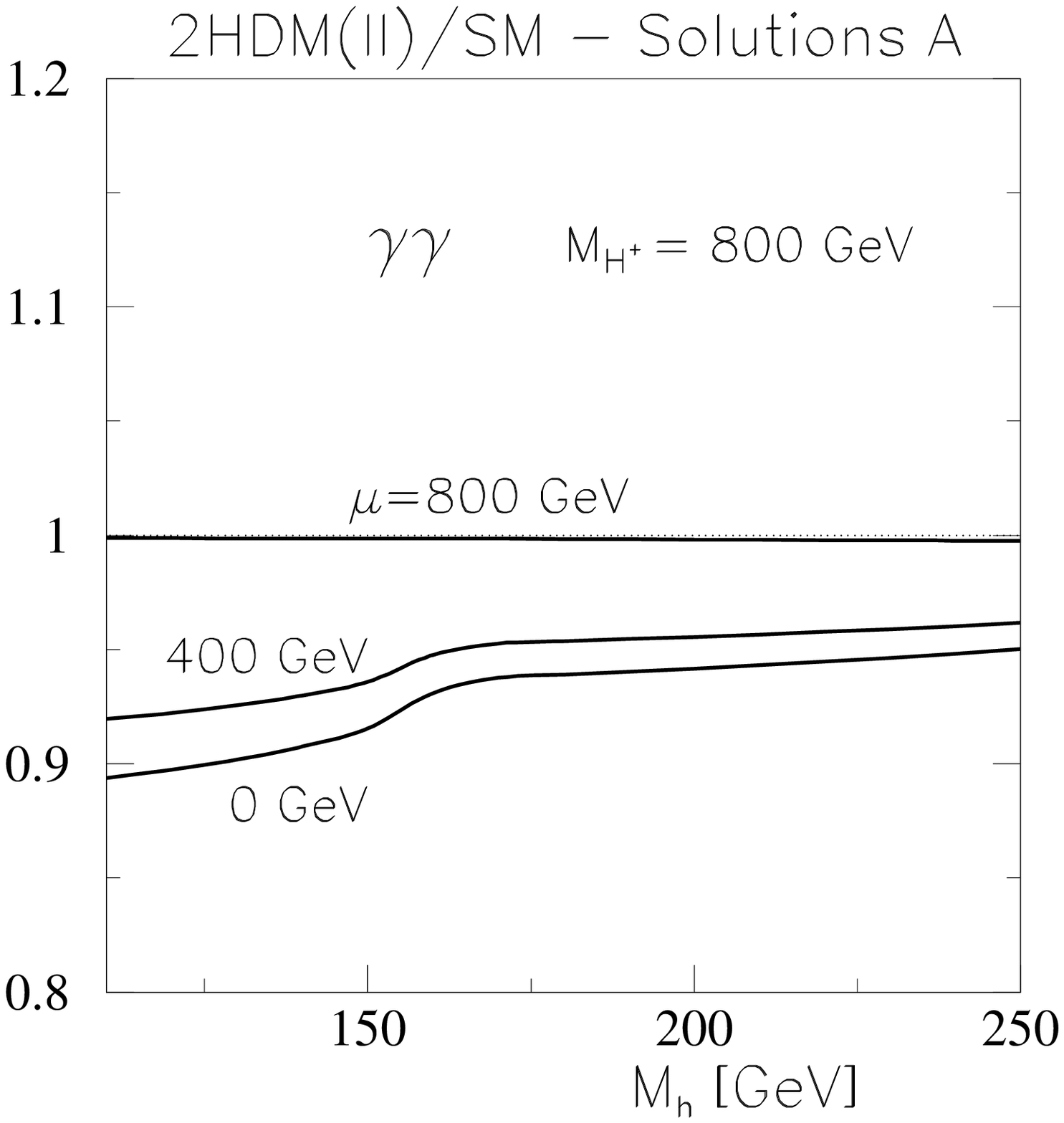}
\includegraphics[height=.3\textheight]{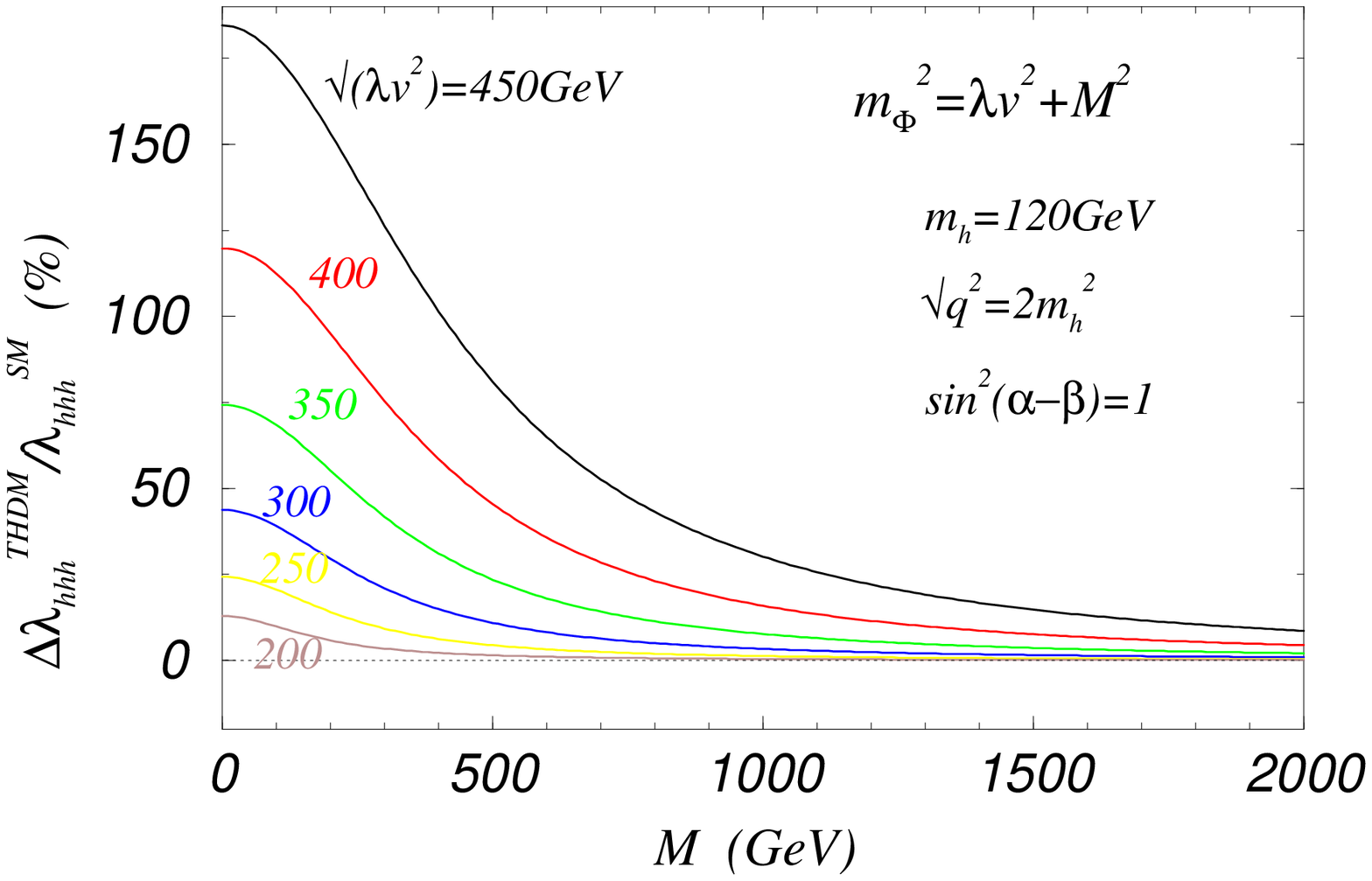}
\caption{Left: Deviation from the SM for $h$ in the $\Gamma_{\gamma \gamma}$ 
due to the 
charged Higgs boson contribution in the SM-like 2HDM II
 for various  $\mu$-values \cite{gko};
Right: Deviation from the SM prediction of a selfcoupling for Higgs boson 
with mass 120 GeV
as a function of mass parameter $M$ (equivalent to $\mu$) 
for various  parameters $\lambda v^2$ \cite{kanemura}.}
\label{fig4}
\end{figure}

It is clear that the Photon Collider  has a large potential in distinguishing 
the SM-like models, what illustrates   Figs.\ref{fig3}. As we mentioned above,
even  in the SM-like limit of the 2HDM II a contribution to $\gamma \gamma h$
coupling due a  heavy  $H^+$  
 leads to a substantial deviation from the SM prediction, i.e. we observe 
 the non-decoupling effect  \cite{gko-ga,gko,hab}. The effects arises  
from $H^+H^-h$ vertex, described by term  proportional to 
$(1-\mu^2/M_{H^+}^2)$, 
where $\mu$ is a (soft-breaking) mass parameter  from the Higgs potential. 
This non-decoupling effect is larger for small   $\mu$, 
and disappears for $\mu=M_{H^+}$, see results presented for a ratio of 
$\Gamma_{\gamma \gamma}$ in 2HDM and SM in 
Fig.~\ref{fig4}(Left).
With an expected precision of the measurement of  $\Gamma_{\gamma \gamma}$,
see e.g. Fig.~\ref{fig3}(Left), such decoupling effect can be seen, and 
moreover one can try to 
constrain the  $\mu$-parameter. The radiative corrections to the $hhh$ 
vertex in the 2HDM II, with the basic couplings to fermions and gauge bosons 
W/Z as in the SM, may also lead to a non-decoupling phenomena, as discussed during this workshop   
\cite{kanemura}. 
For a Higgs-boson mass equal to 120 GeV
the deviation from the SM prediction for the $hhh$ vertex
 can reach even  100 \% 
 (Fig.~\ref{fig4}(Right)).
Note, that in MSSM there is no such effect (decoupling).

\begin{figure}
\hspace*{1cm}\includegraphics[height=.3\textheight]{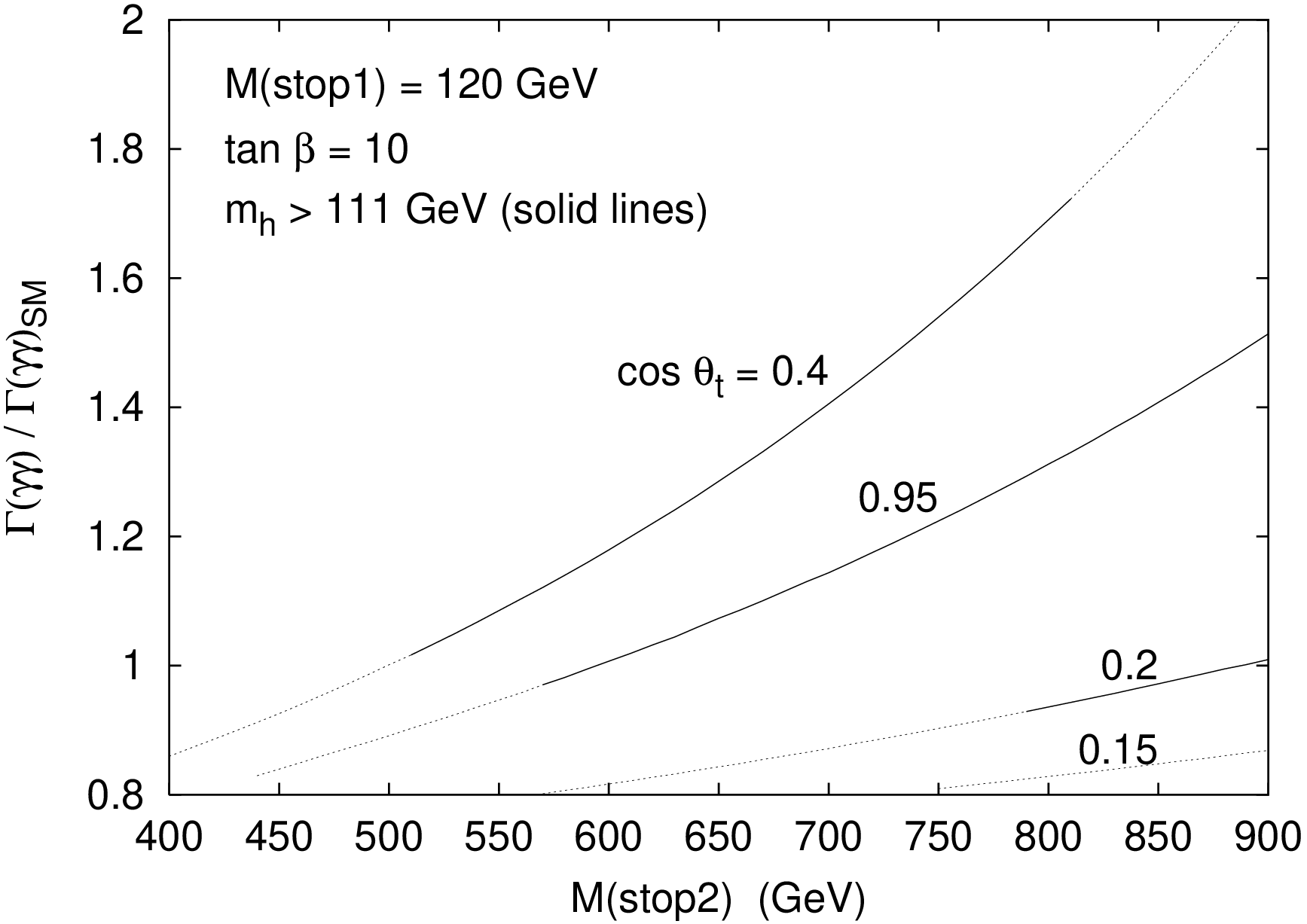}
\includegraphics[height=.3\textheight]{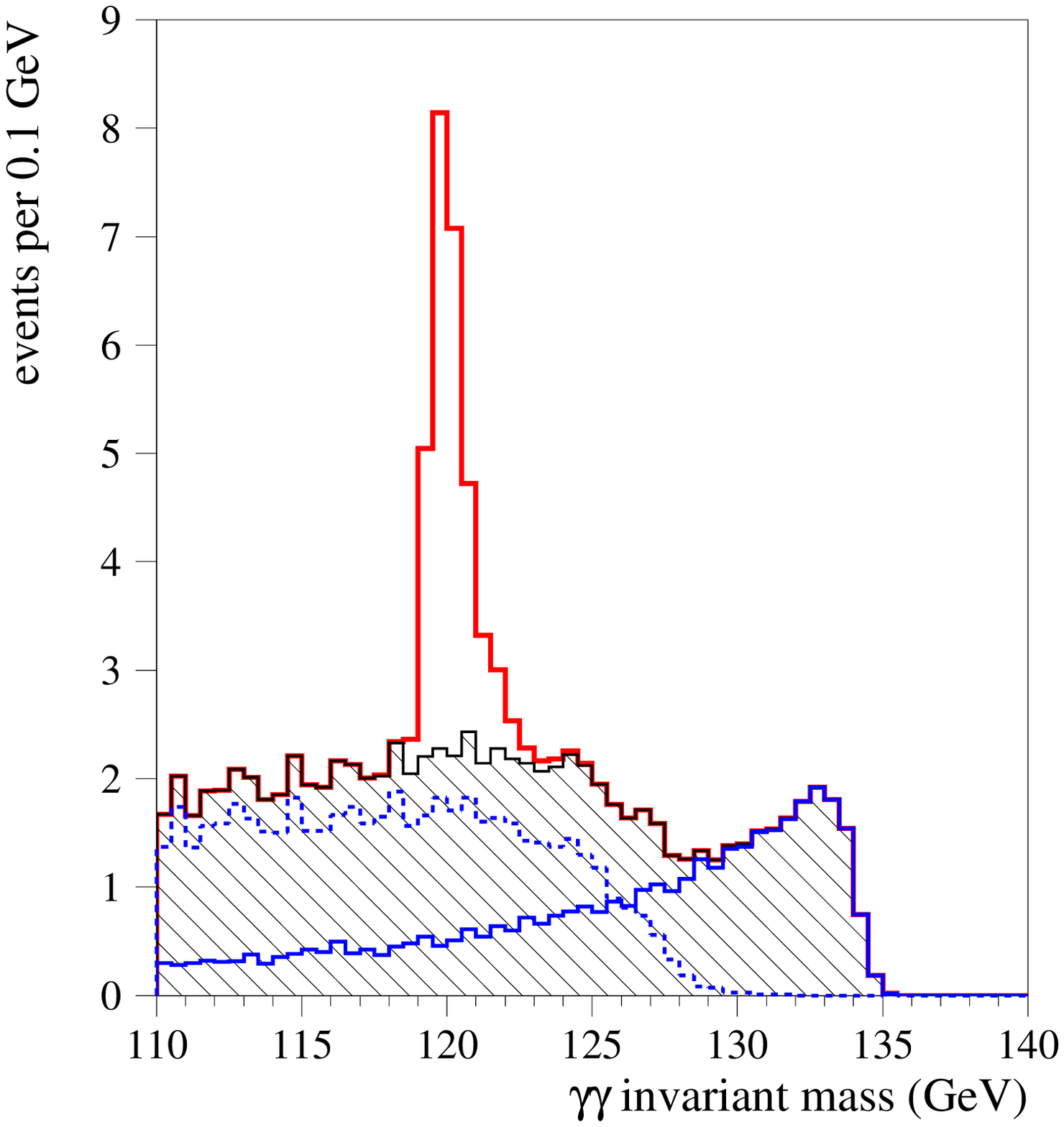}
\caption{Left: Deviation from SM of  the partial width $\Gamma_{\gamma \gamma}$
due to the $\tilde t_2$ as a function of its mass,
for fixed mass of $\tilde t_1$ equal 120 GeV and for various mixing angle
$\cos \theta_t$ \cite{logan}; Right: Number of events of the $\gamma \gamma 
\rightarrow \gamma \gamma$ process with a Higgs resonance at  mass equal to 
120 GeV \cite{schmitt}.}
\label{fig5}
\end{figure}

The precise measurement of  the decay width  $\Gamma_{\gamma \gamma}$ 
can  reveal heavy charged particle circulating in the loop,
e.g. supersymmetric particle. There is for example 
a sensitivity to a  the heavier stop $\tilde {t}_2$ contribution, as discussed
 in \cite{logan}.
 Assuming that the lighter stop $\tilde {t}_1$ and the mixing 
angle $\cos \theta_{\tilde {t}}$ are already known, 
the accuracy of the mass determination is estimated to be 10-20 GeV (for 
500 GeV LC collider and mass of Higgs boson above 110 GeV). Some of  results 
are presented in Fig. \ref{fig5} (Left). There are other possible 
contributions to the $\gamma \gamma h$ loop coupling, for example 
 the (CP-violating)  chargino contribution, see e.g. \cite{bae}. 
The dedicated analysis of such 
contribution in the decoupling regime of MSSM  was performed in 
\cite{Choi:2002rc}, and  we discuss it below.

The production of the SM Higgs-boson with mass equal to 120 GeV
in a process $\gamma \gamma \rightarrow \gamma \gamma$ being
 ``doubly sensitive to $\Gamma_{\gamma \gamma}$'' was analysed in 
\cite{schmitt}. This  analyses leads to an impressive result
 presented in Fig.\ref{fig5} (Right). The real issue  here is a background,
as discussed during the workshop.
 
\subsection{MSSM Higgs particles: A and H}
\begin{figure}
  \includegraphics[height=.3\textheight]{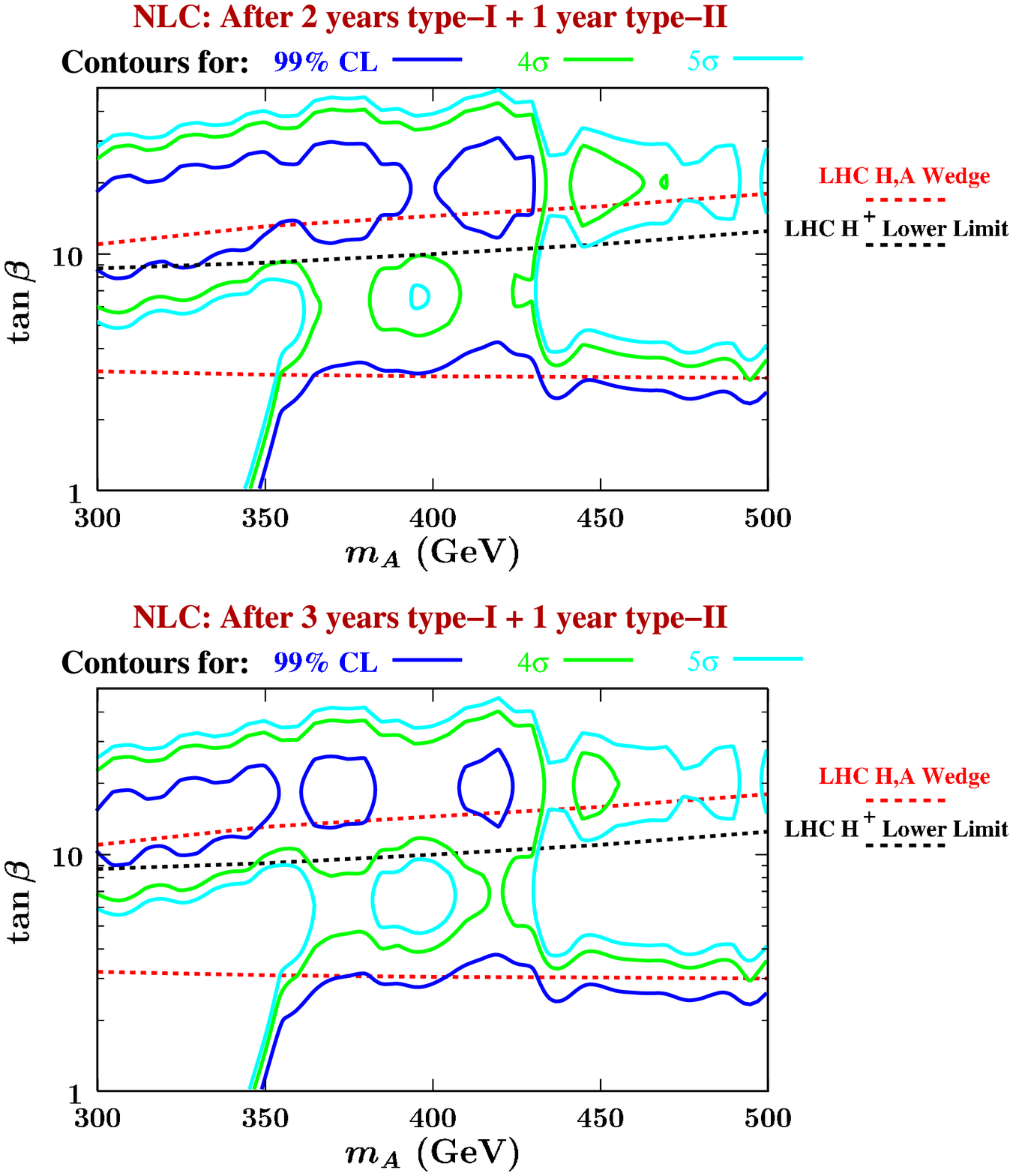}
\includegraphics[height=.3\textheight]{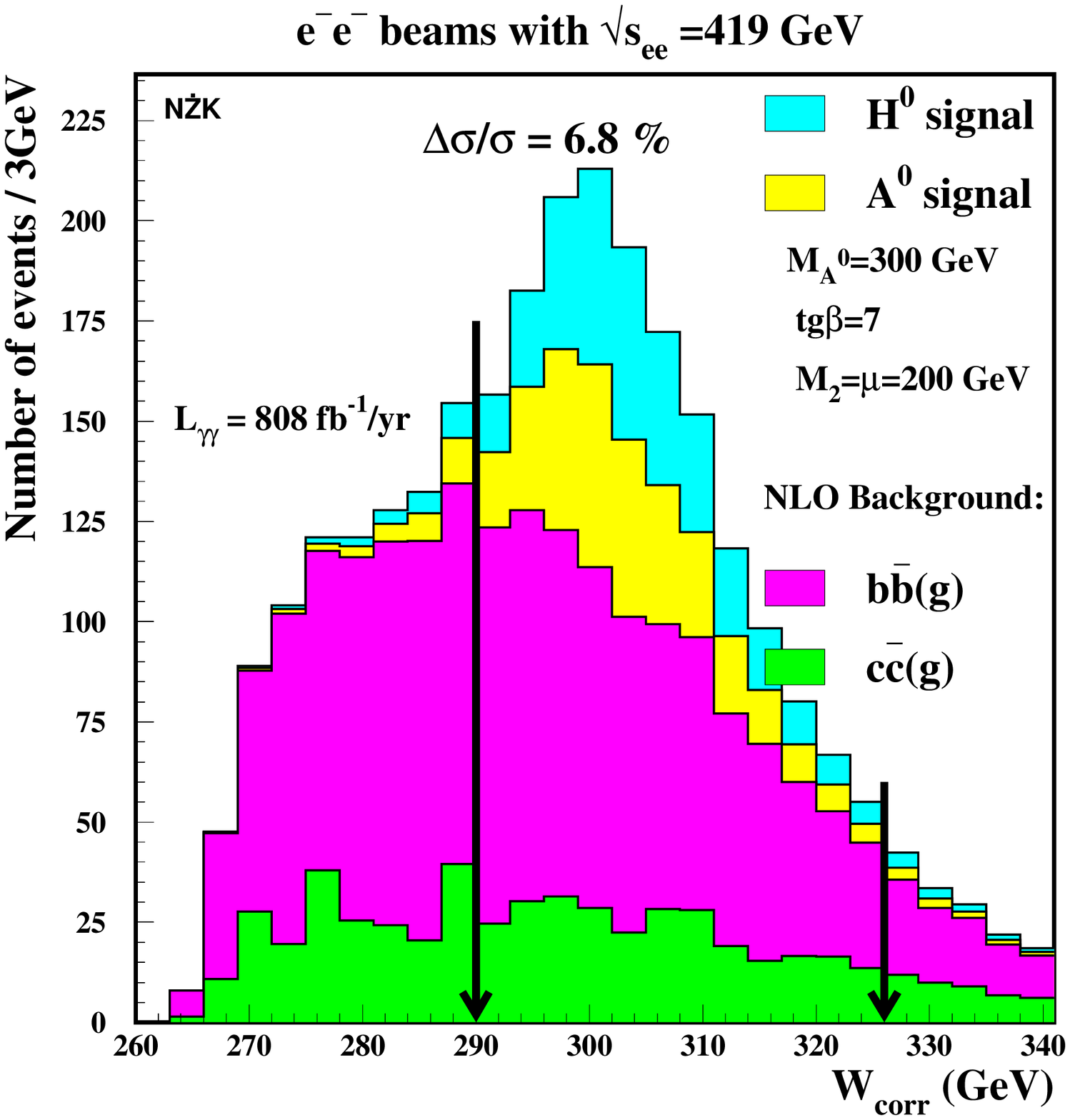}
\caption{Production of  A and H, with  parameters  corresponding to
 the LHC wedge, at the $\gamma \gamma$ collider. 
 Left: Exclusion and discovery limits obtained for NLC collider for 
$\sqrt {ee} =$  630 GeV, after 3 or 4 years of operation (using the  broad and 
peaked spectra) 
\cite{mayda}; Right: Simulation of the H and A signals for TESLA 
collider with energy $\sqrt {ee} =$ 
419 GeV (peaked spectrum) for $M_A=$ 300 GeV, $\tan \beta=7$ \cite{nzk-b-ams}.}
\label{fig6}
\end{figure}

The MSSM Higgs-bosons A and H, in the mass range above 200 GeV and 
with $\tan \beta$ between 6 and 15,
would escape discovery at the LHC (a ``LHC wedge''). On the other hand  they maybe  too heavy 
to be produced 
at the first stage of $e^+e^-$ LC. In this scenario only a light  $h$ 
is expected to 
be observed at future colliders with  couplings to fermions and W/Z 
gauge-bosons as in the SM. At the same time  
 heavy Higgs bosons, A and H, will
decay predominately into $b \bar b$ final state,
and  could be discovered at the Photon Collider, as 
discussed in \cite{mull}, and \cite{asner-gunion,mayda} (Fig. 
\ref{fig6} (Left)). A  new  simulation for TESLA collider
performed in \cite{nzk-b-ams} confirms  these results, see Fig.\ref{fig6} (Right).

A separating of these two heavy Higgs bosons, A and H,
which in the considered scenario are nearly degenerate in mass, 
maybe be very difficult.
According to \cite{mull} this can be done by scanning over an energy 
threshold. 
Other methods of 
separation, in which one  uses information on the CP-properties 
of H and A, are  proposed for the heavier Higgs particles 
decaying into $t \bar t$, 
and  we discuss them  below.
\subsection{Charged Higgs boson}
\begin{figure}
  \includegraphics[height=.3\textheight]{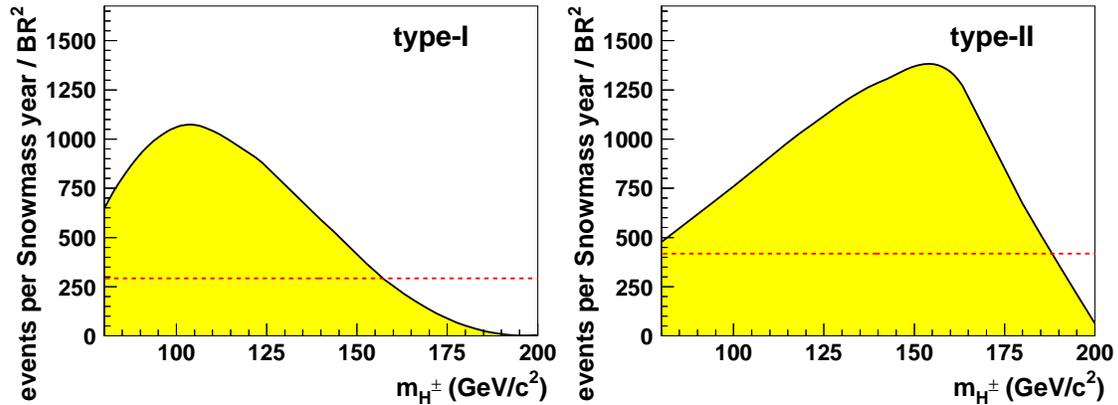}
  \caption{Number of $\gamma
\gamma \rightarrow H^+H^- $ events obtained for a broad and a peaked
 luminosity spectra (type-I and -II, respectively) 
as a function of $M_{H^+}$ \cite{wict}.}
\label{fig7}
\end{figure}

The process   $\gamma \gamma \rightarrow H^+H^-$, within 2HDM II, for the NLC 
version of
the Photon Collider with $\sqrt{s_{ee}}$ = 500 GeV and the product of 
helicity for electron and laser photon  equal to $2 \lambda_e P_c =\pm 0.8 $ 
(the broad (I) and peaked (II) luminosity spectra) were analysed  
in \cite{wict} using
the $\tau^+ \nu_{\tau} \tau^- \bar {\nu_{\tau}}$ final state. 
The background due to $\gamma \gamma \rightarrow W^+W^-$ was taken into
 account, with conclusion that a clean, large signal is obtained up to 
$M_{H^+}\sim $ 195 GeV (Fig. \ref{fig7}). This leads to
3-5 \% uncertainty in determining of the product of the cross section for  
$\gamma \gamma \rightarrow H^+H^-$ and 
 [Br($H^+\rightarrow \tau^+\nu_{\tau}$)]$^2$.
%
\subsection{CP properties of Higgs bosons}

The $\gamma \gamma$ colliders  with the tunable energy and polarization 
(circular or linear) of the  photon-beams can 
be especially  useful for study of the CP properties of the Higgs sector 
\cite{guniongrz,hagiwara}, both to  establish the CP quantum-numbers 
of the  neutral Higgs bosons in the case of a CP-conservation,
and also to find possible effects due to a  CP-violation. 
Note, that the   CP-violating mixing 
between Higgs particles and a simple overlap in mass for heavy neutral
 Higgs bosons in a CP-conserving case  may lead to 
similar effects \cite{Banin:1998ap}.
In testing CP-properties of Higgs particle(s)  one can take 
advantage of  polarization  asymmetries for a signal \cite{guniongrz},
use the interference between the amplitude for production of Higgs 
(signal) and of background,  and get additional  informations  
from the detailed study of the final-state particles. 

If  CP is a good symmetry and Higgs bosons have a definite CP-parity, a
high degree of a linear polarization of photon beams may be  very  useful 
in  distinguishing a relatively light 
CP-even from CP-odd Higgs boson (e.g. H and A in MSSM). This is  
especially important if such Higgs particles are degenerated in  masses, 
as it happends in the MSSM in the scenario with a light SM-like $h$ 
mentioned above, see e.g. Fig.~\ref{fig6}(Right).  
Also, if there are CP-violation effects  leading to a mixing between Higgs 
bosons, linear polarization of the photon beams  looks as an ideal tool.
Still a  circular polarization can  be more useful in searching
and studying very massive Higgs boson, as a transfer of a linear polarization 
from a laser-photon to a photon-beam is 
not efficient for   large $x$, i.e. for $\sqrt{s_{\gamma \gamma}}$ larger than 
half of $\sqrt{s_{ee}}$.
  
To test the CP-property of a Higgs boson one can always use the polarization 
asymmetries \cite{guniongrz}. For heavier masses,
one uses  additional information from the final-state fermions.
For example, identifying Higgs spin and parity in decays to 
ZZ \cite{Choi:2002jk}  can be done  using
 the angular distributions of the  fermions from the Z-boson 
decay, which encode  the helicities of Z's.  
Detailed study was performed for above and below ZZ threshold. 
A realistic  simulation based on this analysis, 
using both the WW and ZZ final-states,
was made recently for the TESLA collider   in \cite{nzk-ams-zz}.

\begin{figure}
  \includegraphics[height=.3\textheight]{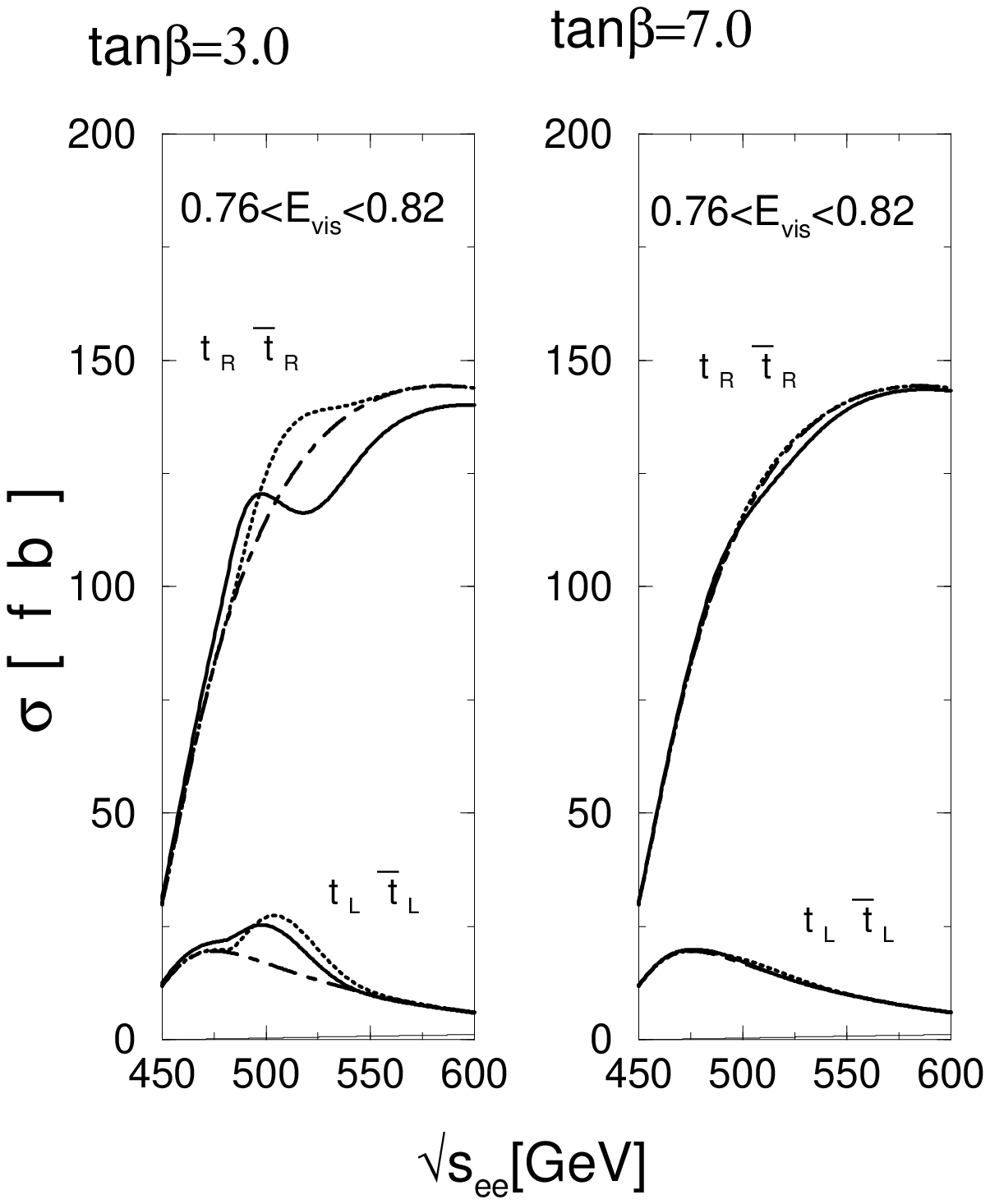}
  \includegraphics[height=.3\textheight]{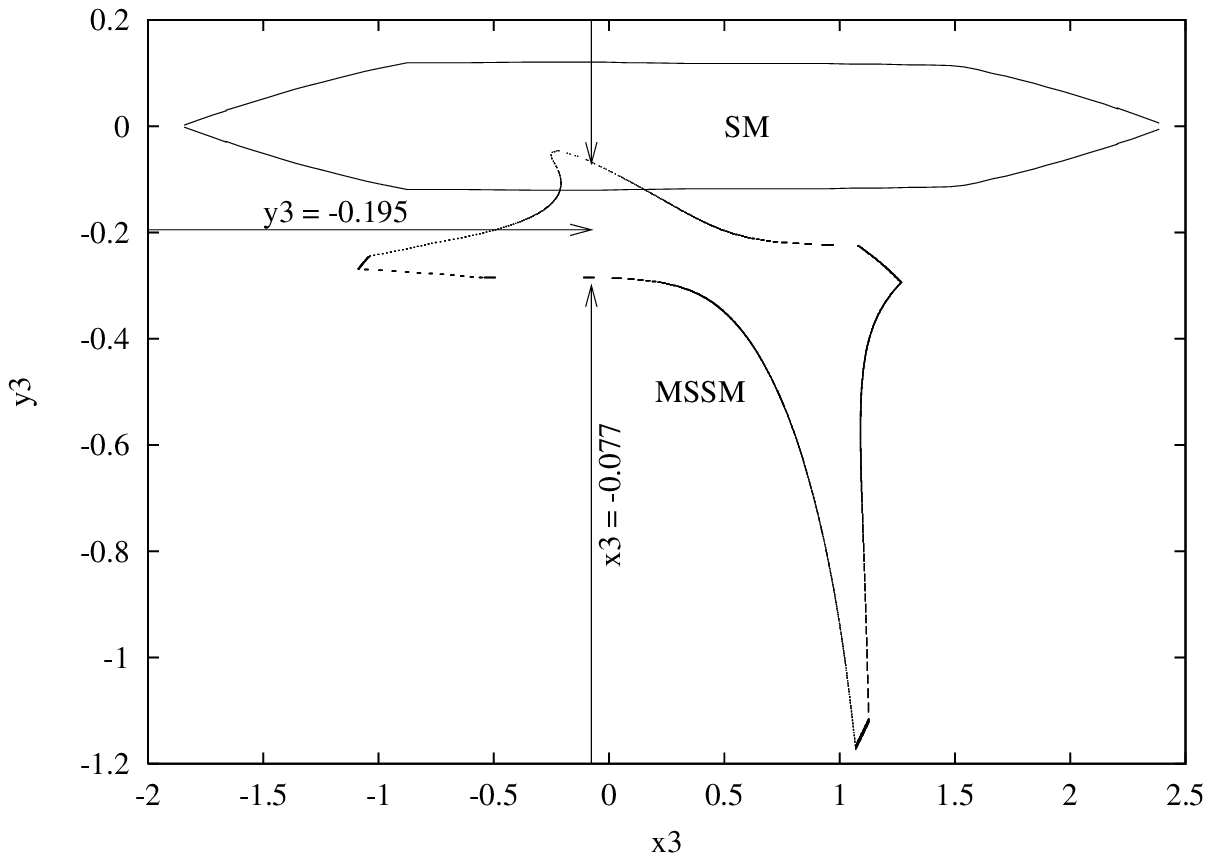}
  \caption{Left:  The inference pattern allowing to distinguish 
CP-odd from CP-even
Higgs particle with mass 400 GeV produced in $\gamma \gamma \rightarrow t \bar t$ \cite{Asakawa:1999gz};
Right: Results from combined analysis of polarization asymmetry and decay lepton charge in  a plane of a CP-even and CP-odd observables, showing a bind regions
for SM and MSSM
 \cite{Godbole:2002qu}. }
\label{fig8}
\end{figure}

For heavier  Higgs bosons 
the decay into a $t \bar t$ final-state can be explored.
The analysis \cite{Asakawa:1999gz} relies on interference between heavy H and 
A, with a small mass gap, and between the Higgs resonances and background for 
a fixed helicities of the decaying $t$ and $\bar t$, 
and it  uses a circular polarization of the photon beams.
This method of distinguishing Higgs bosons and establishing their CP-properties
was found to be  efficient for a small $\tan \beta\sim 3$, see Fig.~\ref{fig8}
(Left). 
A new analysis \cite{a-h} is based  on  interference effects, decay angular 
distribution of $t\bar t$ and in  addition on a phase of the 
$\gamma \gamma \phi $ coupling.
(The importance of measurement of   the phase 
of $\gamma \gamma h $   was pointed out in \cite{nzk-w} for  WW/ZZ channel, 
see Fig.~\ref{fig3}(Right)). This technique allows to test 
 the CP-properties of heavy 
Higgs bosons, what was shown in \cite{a-h}
for a particular case of the CP-conserving  MSSM. 

The CP-violating effects in $\gamma \gamma \rightarrow \phi \ra t\bar t$, 
both in the $\gamma \gamma \phi$ and $t\bar t \phi$ vertices,
were studied in \cite{Asakawa:2000jy}. Model-independent analysis of the 
effects due to Higgs bosons without definite CP-parity was performed
 using the interference among Higgs particles, as well with a background, 
for fixed top-quark helicities.
Photon beams are assumed to be polarized 
(both circular and linear polarization is needed) and  $t$ and $\bar t$ 
helicities are fixed and  equal. The extension of a new analysis reported in 
\cite{a-h} to the CP-violating case is in under way.

 Also in  \cite{Godbole:2002qu} a model-independent analyses was done 
for the same channel.
It was observed that the angular distribution of the decay lepton from 
$t/\bar t$ is independent of any CP violation
in the $tbW$ vertex and hence directly related to  a CP mixing in the
Higgs sector. In the analysis the combined asymmetries
in the initial state (parent) electron  (hence also photon-beam) polarization 
and the final-state lepton-charge have been applied.
If CP-violation is observed, then all the asymmetries, both
for circular and linear 
polarizations, should be used to determine the form factors
describing the  $\gamma \gamma \phi$ and $t\bar t \phi$ vertices.
This method allows to discriminate models, e.g.  SM and MSSM,
 see Fig. \ref{fig8}(Right), where in  a plane of the CP-even and CP-odd 
observables ($x_3$ and $y_3$) the  blind regions for the SM and the 
MSSM are shown.

\begin{figure}
 \includegraphics[height=.3\textheight]{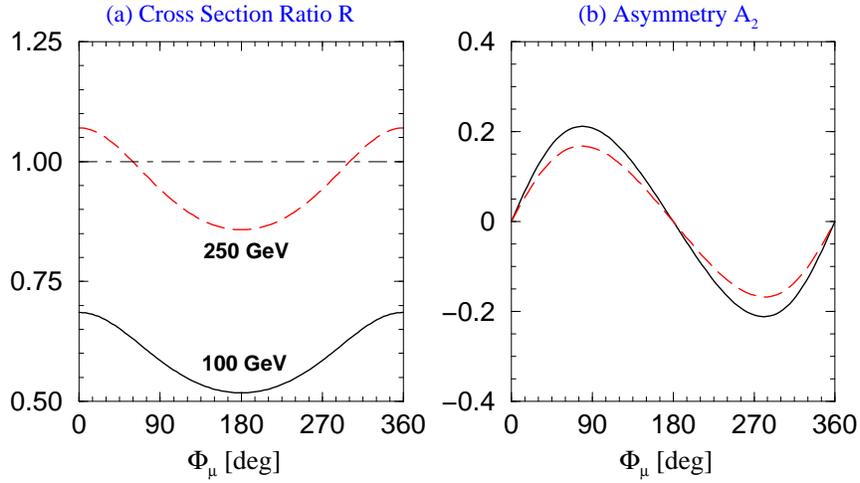}
  \caption{The dependence on the CP-violating phase  $\Phi_\mu$.
Left: the ratio $R$ of the cross section $\hat{\sigma}_0$
          to the SM prediction (150 fb); Right: the CP--odd asymmetry
	  ${\cal A}_2$ for
	  $m_{\tilde{t}_R}=100$ GeV (solid lines) and
	  $m_{\tilde{t}_R}=250$ GeV (dashed lines), respectively 
\cite{Choi:2002rc}.}
\label{fig9}
\end{figure}

In the CP-violating SUSY, the spin-zero top squarks contribute only to the 
CP-even part of the $h\gamma\gamma$ coupling,
while charginos can contribute to the CP-odd as well as the CP-even
parts \cite{bae}, leading to a CP-violation.
It was  demonstrated in \cite{Choi:2002rc},
that the measurement of the lightest Higgs boson, produced
in the collision of the linearly-polarized photon-beams, allows  to confirm the
existence of the CP-violating chargino contributions to the $\gamma \gamma h$
coupling { even in the decoupling regime} of MSSM. Results were 
obtained in a specific  CP-violating scenario, in agreement with existing
constraints. The obtained  predictions depend strongly on  
the CP-violating phase $\Phi_{\mu}$ both  for a ratio of 
cross section to the SM  presented in Figs.~\ref{fig9}(Left), 
and  for the polarization asymmetry $A_2$, see  Figs.~\ref{fig9}(Right).

%

\section{Summary}
The Photon Collider,
 both in the \gaga \, and \ega \, modes  
can  provide  valuable informations, some of them not accessible at the   
$e^+e^-$ colliders, nor at Tevatron or LHC. 
In light of realistic simulations of various golden processes presented 
during this workshop and as result of a special  panel discussion 
on a need of a Photon Collider  
a ``little consensus'' was reached:  
a Photon Collider should be planned  as an option at each $e^+e^-$ LC
project.


\begin{theacknowledgments}
I would like to thank the Organizing Commitee for this excellent meeting
and all conveners and speakers for interesting joint sessions.  
I am especially grateful to Tohru Tokahashi for a nice collaboration,
and also to Jeff Gronberg. 
Work supported  in part by Polish Committee for Scientific Research, Grant
2~P~03~B~05119 (2002-2003), 5 P03B 121 20 (2002-2003), and by the European 
Community's Human Potential Programme under contract HPRN-CT-2000-00149 Physics
at Colliders (2002) and EURIDICE (2003).
\end{theacknowledgments}


\begin{thebibliography}{9999}
\bibitem{ginz-princ}
I.~F. Ginzburg, G.~L. Kotkin, V.~G. Serbo and V.~I. Telnov,{ Pizma~ZhETF\/} 
{\bf 34} (1981) 514,({JETP Lett.} {\bf 34} (1982) 491),
 Nucl.~Instrum.~Meth.~A {\bf 205} (1983) 47;
I.~F. Ginzburg, G.~L. Kotkin, S.~L. Panfil, V.~G. Serbo and V.~I. Telnov,
 { Nucl.~Instrum.~Meth.\/} A {\bf 219} (1984) 5;
\bibitem{telnov}V.~Telnov, Nucl.~Instrum.~Meth. A {\bf 294} (1990) 72 and  
A {\bf 355} (1995) 3, A {\bf 472} (2001) 43 (hep-ex/0010033),
 A {\bf 494} (2002) 35
[arXiv:hep-ex/0207093].
\bibitem{space} Linear Collider Report from WW Study Group: Understanding 
`` Matter, Energy, Space and Time''  
\bibitem{gg2000}Proc. of Int. Workshop on High Energy Photon Colliders, 
Hamburg, April 2000;  
Nucl.\ Instrum.\ Meth.\ A {\bf 472} (2001)
\bibitem{badelek}
B.~Badelek {\it et al.}  [ECFA/DESY Photon Collider Working Group
                  Collaboration],
arXiv:hep-ex/0108012.
\bibitem{acfa}
S.~Kiyoura, S.~Kanemura, K.~Odagiri, Y.~Okada, E.~Senaha, S.~Yamashita and Y.~Yasui,
arXiv:hep-ph/0301172; 
I.~Watanabe {\it et al.},
KEK-REPORT-97-17.
\bibitem{tohru}
T.~Takahashi,
Nucl.\ Instrum.\ Meth.\ A {\bf 472} (2001) 4.


\bibitem{bbb}
M.~Baillargeon, G.~Belanger and F.~Boudjema,
colliders,''
Phys.\ Rev.\ D {\bf 51} (1995) 4712
[arXiv:hep-ph/9409263] and 
arXiv:hep-ph/9405359.

\bibitem{golden}
S.~J.~Brodsky and P.~M.~Zerwas,
Nucl.\ Instrum.\ Meth.\ A {\bf 355} (1995) 19
[arXiv:hep-ph/9407362].
E.~Boos {\it et al.},
Nucl.\ Instrum.\ Meth.\ A {\bf 472} (2001) 100
[arXiv:hep-ph/0103090].
\bibitem{hagiwara}
K.~Hagiwara,
Nucl.\ Instrum.\ Meth.\ A {\bf 472} (2001) 12
[arXiv:hep-ph/0011360].

\bibitem{mayda}
D.~Asner, B.~Grzadkowski, J.~F.~Gunion, H.~E.~Logan, V.~Martin, M.~Schmitt 
and M.~M.~Velasco,
arXiv:hep-ph/0208219.


\bibitem{asner-gunion}
D.~M.~Asner, J.~B.~Gronberg and J.~F.~Gunion,
Phys.\ Rev.\ D {\bf 67} (2003) 035009
[arXiv:hep-ph/0110320].
\bibitem{cainphocolcircecompaz}
{\bf CAIN}
P.~Chen, G.~Horton-Smith, T.~Ohgaki, A.~W.~Weidemann and K.~Yokoya,
Nucl.\ Instrum.\ Meth.\ A {\bf 355} (1995) 107;
{\bf PHOCOL} V.~Telnov, \textit{A code PHOCOL for the simulation of 
luminosities
and backgrounds at photon colliders};
{\bf CIRCE}
 T.~Ohl,
Comput.\ Phys.\ Commun.\  {\bf 101} (1997) 269
[arXiv:hep-ph/9607454]; 
{\bf COMPAZ}
A.~F.~Zarnecki,
Acta Phys.\ Polon.\ B {\bf 34} (2003) 2741
[arXiv:hep-ex/0207021].
\bibitem{MC}
{\bf HERWIG 6.5}, G. Corcella, I.G. Knowles, G. Marchesini, S. Moretti, 
K. Odagiri, P. Richardson, M.H. Seymour and B.R. Webber, 
JHEP 0101 (2001) 010 [hep-ph/0011363]; hep-ph/0210213;
{\bf PYTHIA} T. Sj\"ostrand, P. Ed\'en, C. Friberg, L. L\"onnblad, G. Miu, 
S. Mrenna and 
E. Norrbin,Computer Phys. Commun. 135 (2001) 238 (hep-ph/0010017);
{\bf PANDORA} 
http://www-sldnt.slac.stanford.edu/nld/new/Docs/Generators/PANDORA.htm; 
{\bf GRACE} F.~Yuasa {\it et al.},
Prog.\ Theor.\ Phys.\ Suppl.\  {\bf 138} (2000) 18
[arXiv:hep-ph/0007053];
{\bf CompHEP}
 A.Pukhov, E.Boos, M.Dubinin, V.Edneral, V.Ilyin, D.Kovalenko, A.Kryukov, 
V.Savrin, S.Shichanin, and A.Semenov.
Preprint INP MSU 98-41/542, hep-ph/9908288. 

\bibitem{frank} F. Kraus, talk at LC workshop (St. Malo, 2002).
\bibitem{cjkl}
F.~Cornet, P.~Jankowski, M.~Krawczyk and A.~Lorca,
arXiv:hep-ph/0212160, 
presented at the workshop.
\bibitem{adr}
R.~M.~Godbole, A.~De Roeck, A.~Grau and G.~Pancheri,
arXiv:hep-ph/0305071, and this proceedings.
\bibitem{shifman}
A.~I.~Vainshtein, V.~I.~Zakharov and M.~A.~Shifman,
Sov.\ Phys.\ Usp.\  {\bf 23} (1980) 429
[Usp.\ Fiz.\ Nauk {\bf 131} (1980) 537].

\bibitem{lan} L.~D.~Landau, Doklady Acad. Nauk CCCP, tom LX, N02, 207, 1948;
C.~N.~Yang,
Phys.\ Rev.\  {\bf 77} (1950) 242.


\bibitem{tohru96}
T.~Ohgaki, T.~Takahashi, I.~Watanabe and T.~Tauchi,
Int.\ J.\ Mod.\ Phys.\ A {\bf 13} (1998) 2411.
T.~Ohgaki, T.~Takahashi and I.~Watanabe,
Phys.\ Rev.\ D {\bf 56} (1997) 1723
[arXiv:hep-ph/9703301].

\bibitem{ssr}
G.~Jikia and S.~S\"oldner-Rembold, Nucl.~Instrum.~Meth.~A {\bf 472} (2001) 133, hep-ex/0101056.
\bibitem{nzk-b}
P.~Nie\.zurawski, A.~F.~\.Zarnecki and M.~Krawczyk,
photon collider at TESLA,''
Acta Phys.\ Polon.\ B {\bf 34} (2003) 177
[arXiv:hep-ph/0208234],  and this proceedings. 
\bibitem{nzk-b-ams} 
P.~Nie\.zurawski, A.~F.~\.Zarnecki and M.~Krawczyk,
LC workshop, April 2003,  Amsterdam, 
``New results for $\gamma \gamma \rightarrow H \rightarrow b \bar b$ 
in SM and MSSM''.
\bibitem{moenig} K. Moenig, this proceedings.
\bibitem{rindani}S. Rindani, pleanary talk.
\bibitem{godfrey}
S.~Godfrey and M.~A.~Doncheski,
Phys.\ Rev.\ D {\bf 65} (2002) 015005 
[arXiv:hep-ph/0108268].
\bibitem{espiru}
P.~Ciafaloni and D.~Espriu,
Phys.\ Rev.\ D {\bf 56} (1997) 1752
[arXiv:hep-ph/9612383];
\bibitem{gko-ga}
I.~F.~Ginzburg, M.~Krawczyk and P.~Osland,
arXiv:hep-ph/9909455,
arXiv:hep-ph/0101331,
Nucl.\ Instrum.\ Meth.\ A {\bf 472} (2001) 149
[arXiv:hep-ph/0101229],
arXiv:hep-ph/0101208.
\bibitem{gko}
I.~F.~Ginzburg, M.~Krawczyk and P.~Osland,
arXiv:hep-ph/0211371,
 in this proceedings.
\bibitem{hab}
J.~F.~Gunion and H.~E.~Haber,
Phys.\ Rev.\ D {\bf 67} (2003) 075019
[arXiv:hep-ph/0207010].

\bibitem{cidwaj}
D.~A.~Morris, T.~N.~Truong and D.~Zappala,
Phys.\ Lett.\ B {\bf 323} (1994) 421
[arXiv:hep-ph/9310244],
I.~F.~Ginzburg and I.~P.~Ivanov,
Phys.\ Lett.\ B {\bf 408} (1997) 325
[arXiv:hep-ph/9704220].


\bibitem{nzk-w}
P.~Nie\.zurawski, A.~F.~\.Zarnecki and M.~Krawczyk,
JHEP {\bf 0211} (2002) 034[arXiv:hep-ph/0207294], and  this proceedings.
\bibitem{kanemura}
S.~Kanemura, S.~Kiyoura, Y.~Okada, E.~Senaha and C.~P.~Yuan,
Phys.\ Lett.\ B {\bf 558} (2003) 157
[arXiv:hep-ph/0211308].
and this proceedings.

\bibitem{logan}H. Logan, presented at the workshop.
\bibitem{bae}
S.~Bae, B.~Chung and P.~Ko,

arXiv:hep-ph/0205212.
\bibitem{Choi:2002rc}
S.~Y.~Choi, B.~c.~Chung, P.~Ko and J.~S.~Lee,
Phys.\ Rev.\ D {\bf 66} (2002) 016009
[arXiv:hep-ph/0206025], presented at the workshop.
\bibitem{schmitt}M. Schmitt, presented at the workshop
\bibitem{mull}
M.~M.~Muhlleitner, M.~Kramer, M.~Spira and P.~M.~Zerwas,
Phys.\ Lett.\ B {\bf 508} (2001) 311
[arXiv:hep-ph/0101083].
\bibitem{wict}V. Martin, presented at the workshop.
\bibitem{guniongrz}
B.~Grzadkowski and J.~F.~Gunion,
Phys.\ Lett.\ B {\bf 294} (1992) 361
[arXiv:hep-ph/9206262],
M.~Kramer, J.~H.~Kuhn, M.~L.~Stong and P.~M.~Zerwas,
Z.\ Phys.\ C {\bf 64} (1994) 21,
W.~G.~Ma, C.~H.~Chang, X.~Q.~Li, Z.~H.~Yu and L.~Han,
Commun.\ Theor.\ Phys.\  {\bf 26} (1996) 455,
Commun.\ Theor.\ Phys.\  {\bf 27} (1997) 101,
G.~J.~Gounaris and G.~P.~Tsirigoti,
Phys.\ Rev.\ D {\bf 56}  (1997) 3030
[Erratum-ibid.\ D {\bf 58} (1998)  059901]
[arXiv:hep-ph/9703446].
\bibitem{Banin:1998ap}
A.~T.~Banin, I.~F.~Ginzburg and I.~P.~Ivanov,
Phys.\ Rev.\ D {\bf 59} (1999) 115001
[arXiv:hep-ph/9806515].
I.~F.~Ginzburg and I.~P.~Ivanov,
Eur.\ Phys.\ J.\ C {\bf 22} (2001) 411
[arXiv:hep-ph/0004069].
\bibitem{Choi:2002jk}
S.~Y.~Choi, D.~J.~Miller, M.~M.~Muhlleitner and P.~M.~Zerwas,
Phys.\ Lett.\ B {\bf 553} (2003) 61
[arXiv:hep-ph/0210077].
\bibitem{nzk-ams-zz} P. Nie\.zurawski, A.F. \.Zarnecki, M. Krawczyk,
LC Workshop, April 2003,   Amsterdam,
``Measurement of angular distributions for
$\gamma \gamma \rightarrow h \rightarrow ZZ/WW \rightarrow  ll jj \;/ \; 4j $''.
\bibitem{Asakawa:1999gz}
E.~Asakawa, J.~i.~Kamoshita, A.~Sugamoto and I.~Watanabe,
Eur.\ Phys.\ J.\ C {\bf 14} (2000) 335
[arXiv:hep-ph/9912373].
\bibitem{a-h}
E.~Asakawa and K.~Hagiwara,
arXiv:hep-ph/0305323.
\bibitem{Asakawa:2000jy}
E.~Asakawa, S.~Y.~Choi, K.~Hagiwara and J.~S.~Lee,
Phys.\ Rev.\ D {\bf 62} (2000) 115005
[arXiv:hep-ph/0005313].
\bibitem{Godbole:2002qu}
R.~M.~Godbole, S.~D.~Rindani and R.~K.~Singh,
Phys.\ Rev.\ D {\bf 67} (2003) 095009
[arXiv:hep-ph/0211136], and in this proceedings.
 \end{thebibliography}
\end{document}